\documentstyle[prl,aps,twocolumn]{revtex}
\begin{document}
\twocolumn[\hsize\textwidth\columnwidth\hsize\csname
@twocolumnfalse\endcsname

\draft

\title{Multiparticle entanglement and its applications to cryptography}
\author{Julia Kempe \footnotemark[1]}
\address{Department of Mathematics, University of California, Berkeley \footnotemark[2]}
\address{\'Ecole Nationale Superieure des T\'el\'ecommunications, Paris, France}

\date{\today}
\maketitle

\pagestyle{plain}
\pagenumbering{arabic}

\begin{abstract}

Entanglement between three or more parties exhibits a realm of
properties unknown to two-party states. Bipartite states are easily classified using the {\em Schmidt decomposition}. The Schmidt coefficients of a
bipartite pure state encompass all the non-local properties of the
state and can be "seen" by looking at one party's density matrix
only. \newline Pure states of three
and more parties however lack such a simple form. They have more invariants under local unitary
transformations than any one party can "see" on their sub-system.
These "hidden non-localities" will allow us to exhibit a class of
multipartite states that cannot be distinguished from each other by
any party. Generalizing a result of BPRST$^1$ and
using a recent result by Nielsen we will show that these states
cannot be transformed into each other by local actions and
classical communication. Furthermore we will use an orthogonal
subset of such states to hint at applications to cryptography and illustrate an
extension to quantum secret sharing (using recently suggested $((n,k))$-threshold schemes).
\end{abstract}
\hspace{1cm} ] \narrowtext

\footnotetext{$^1$ Bennett, Popescu, Rohrlich, Smolin and Thapliyal}
\footnotetext{\footnotemark[1] Email: kempe@math.berkeley.edu}
\footnotetext{\footnotemark[2] visiting California Institute of Technology, Pasadena}
\section{Introduction}

The entanglement properties of {\em bipartite pure states} have
already been treated extensively. The analysis of entanglement and its
properties for these states is much easier than for $3$ or more party shared
states due to a particularly convenient form that captures all
non-local parameters: the (unique) {\em Schmidt decomposition} \cite{Peres}.\newline
An interesting question which arises in attempts to classify
entanglement is which states can be obtained from a given state
if we allow {\em local actions and classical communication} of
the parties. By classical communication we mean an a priori unlimited amount of two-way classical communications. We will call these transformations of a $k$-party state
{\em $k$-LOCC} ($k$-party local operations and classical
communication). The crucial difference between pure local unitary
action and LOCC is that each party may perform (generalized)
measurements on its subsystem and broadcast the outcomes via
classical channels between the parties. The other parties may
choose their subsequent actions conditional on the outcomes of
these measurements.\newline
For {\em bipartite} pure states Nielsen \cite{Michael} has recently found necessary and sufficient
conditions  for the process of entanglement transformation via $2$-LOCC
to be possible. A key tool in this result is the Schmidt-decomposition
of bipartite states and the conditions involve the Schmidt-coefficients of the states only. \newline
So once we are given the density matrix of one party a bipartite pure state contains no more secret to us: The  
 eigenvalues of one party's density matrix completely characterize the
 state (up to equivalence under local unitary operations) and give us
 complete knowledge about its entanglement transformation properties under local
 operations and classical communication between the parties. In other words given a sufficient supply of copies of a certain state shared by two parties each of the parties is able to determine (up to a certain precision) its equivalence class under local unitaries {\em and} which other states it can be transformed into via $2$-LOCC.\newline
The situation is drastically different for {\em multipartite} states involving more than two parties. No convenient (locally invariant) form--analogous to the Schmidt-decomposition--can be given. The number of invariants of a state under local unitaries grows exponentially with the number of parties (see Section \ref{hidden}). Attempts to find
canonical points on the orbits of multipartite states have been
made \cite{LindenPopescuSudbery}, but yield unwieldy outcomes. We will say that two multiparty-states are {\em unitarily equivalent}
($|\Psi \rangle \sim |\Phi \rangle $) if they can be transformed
into each other by local (single-party) unitary operations only
(without classical communication). Linden
and Popescu \cite{LindenPopescu97} have given a lower bound on the
number of parameters needed to describe equivalence classes of
multipartite states. To parameterize inequivalent states they also exhibited an explicit polynomial form for 
invariants of a multipartite state under local unitaries (see
Section \ref{incom}).  Some of these invariants are functions of the eigenvalues of the local density matrices of all parties. For three (and more) parties however the number of
independent invariants under local unitaries is bigger than the number
of independent eigenvalues of all local density matrices. This means that if we get all possible information from
each party's subsystem there will be invariants under local unitaries
that we cannot determine. We will call these parameters
{\em hidden non-localities} of our quantum-state. \newline Complete
knowledge of each local system thus does not give us complete information
on the equivalence class of the multipartite state under local unitary operations.\newline
Let's review Nielsen's result to make the difference between bipartite
and multipartite states more precise: For bipartite pure states and $2$-LOCC
there is a partial ordering on the states that characterizes their
mutual entanglement transformation properties \cite{Michael}:
\begin{equation}
|\Psi \rangle \stackrel{2-\text{LOCC}}{\longrightarrow }|\Phi \rangle
\,\,\,\,\text{iff}\,\,\,\,\rho _{A}^{\Psi }\prec \rho _{A}^{\Phi }
\label{majo}
\end{equation}
where $\rho _{A}$ is the density matrix of one party and $\rho
_{A}^{\Psi }\prec \rho _{A}^{\Phi }\ $means that the eigenvalues $\lambda_1^{\Psi},\ldots,\lambda_k^{\Psi}$
of $\rho _{A}^{\Psi }$ \ are majorized by the eigenvalues $\lambda_1^{\Phi},\ldots,\lambda_k^{\Phi}$ of $\rho
_{A}^{\Phi }$ , i.e.
\begin{equation}
\sum_{i=1}^{k}\lambda _{i}^{\Psi \downarrow }\leq
\sum_{i=1}^{k}\lambda _{i}^{\Phi \downarrow }\;\;\;\;\;\;\forall k
\end{equation}
The arrow indicates that the eigenvalues have to be
put into decreasing order.\newline This gives a partial ordering
in the space of all non-local parameters of bipartite states
(remember that the non-local parameters are the independent eigenvalues of the density matrix
of one party). In the case of just two qubits shared by two
parties this even gives a {\em total ordering} on the states
meaning that given two states either the first can be transformed
into the second or vice versa (there is only one independent Schmidt-coefficient).  Among {\em higher dimensional bipartite states} 
however we also find sets of states that cannot be transformed into 
each other either way by LOCC. These states have been termed
{\em incommensurate}.  The smallest system to provide us with two bipartite incommensurate pure states is the $9$-dimensional space of two {\em qutrits} \cite{Michael}. Also note that the commensurateness or
incommensurateness of two bipartite states can be immediately
identified by looking at the density matrix of one subsystem. 
\newline
Two bipartite pure states whose one-party density matrices have the same eigenvalues are {\em always} mutually obtainable from each other via $2$-LOCC. Also
\begin{equation}
\label{id}
\frac{1}{n} I\prec \rho_A^{\Phi} \;\;\;\;\;\;\; \forall \; |\Phi\rangle
\end{equation}
implies that starting with an EPR-type bipartite state (unique up to local unitaries with the property that its density matrix obtained by tracing out one party is proportional to the identity matrix) we can extract {\em every} given bipartite state $|\Phi\rangle$ with local operations and classical communication. The partially ordered set of states under $2$-LOCC has just {\em one} maximal state (up to unitary equivalence)!\newline
We will show that this structure is very different for
{\em multipartite states}. BPRST$^1$ \cite{BPRST}, \cite{Popescupc} have found
two $3$-party states--each party having $2$ qubits--of dimension
$2^{6}$ that are incommensurate although all of their sub-density
matrices are identical.
 Following their argument we will use Nielsen's result (\ref{majo}) to generalize
their proof and show that even for the smallest $3$-partite state (of
dimension $8$) there are incommensurate states that have {\em
  identical or similar} local density matrices. Their incommensurateness can
not be ``seen'' by looking at subsystems of the state (it is
{\em hidden}). We connect {\em hidden
  non-localities} to {\em hidden incommensurateness} to see that two multipartite states with similar density matrices on each party are incommensurate if and only if they are not unitarily equivalent.\newline
We give some examples of {\em locally equivalent $k$-LOCC incommensurate states}. We will suggest how to ``encode into hidden non-localities'' with the help of an orthogonal subset of such states. These states have the property that they are totally indistinguishable from each other for each party alone and cannot be transformed into each other by local operations and classical communication between the parties. Furthermore we can find a set of such states that are  {\em maximal} in the sense that they cannot be obtained from any other (unitarily not equivalent) state by $k$-LOCC. Only if the parties perform a collective (orthogonal) measurement they will be able to (perfectly) distinguish these states. This area needs further exploration.\newline
We will analyze a recently suggested cryptographic protocol \cite{CGL}
for {\em quantum secret sharing} to identify a class of incommensurate
and locally equivalent states in them. An $((n,k))$ threshold scheme
($k < n$) is a method to encode and divide a secret quantum state
between $n$ parties such that from any $k$ shares the state can be perfectly recovered and from any $k-1$ or fewer shares no information whatsoever about the state can be inferred.\newline
The scheme as introduced in \cite{CGL} assumes that all parties are honest when they participate in recinstructing the secret. Allowing for the possibility of some parties being dishonest in order to retrieve the secret alone we will show how the scheme can be ``misused'' for cheating by one party if it is used to encode a ``classical'' bit and how this cheating can be prevented by using incommensurate locally identical states.

\section{Counting Hidden Non-Localities}
\label{hidden}
 Linden and Popescu \cite{LindenPopescu97} have classified the orbits of
multipartite states under local unitary operations and determined
the dimension of generic orbits and the number of parameters
needed to describe the location of such an orbit in Hilbert
space.\\ In the case of $k$ parties each having one spin 1/2
particle (qubit) there at least $ 2^{k+1}-2-3k$ real parameters that
characterize non-local properties.  (Initially each
$2^{k}$-dimensional state has $2^{k}$ complex parameters and the
requirement of unit norm leaves $2^{k+1}-1$ real parameters. The
group of equivalence transformations is $U(1)\times SU(2)\times
SU(2)\times \ldots \times SU(2)$--each local unitary $U(2)\simeq
U(1)\times SU(2)$ but each local phase can be factored out to one
single global phase. The group (and the generic orbit) then has
dimension $3k+1$ or less.)\\ Furthermore an explicit form for polynomial
invariants of an orbit has been given \cite{LindenPopescu97} (see
Section \ref{incom}). To get a picture of how the number of
{\em hidden
  non-localities} grows let's analyze the three, four and five-party
  cases. We can easily count parameters in the three party spin 1/2-case: We
have $5$ (independent) non-local parameters but only three (in
general) different density matrices each characterized by one
eigenvalue. So there are $2$ non-local parameters that we cannot
``see'' by only looking at various subsystems of our entangled
state. Now let's look at the four-party spin-1/2 case. Here we
have at least $18$ non-local parameters but at most $7$ independent
sub-density matrices where two density matrices are {\em independent}
if the eigenvalues of the first do not completely determine the
eigenvalues of the second. The four one-party matrices each have one
non-local parameter and the $3$ independent density matrices of $2$
joint parties have at most $3$ parameters each thus leaving a total of
at least $18-4\ast 1-3\ast 3=5$ hidden non-localities. Of the $18$
non-local parameters $14$ cannot be seen by only one party alone, they are {\em hidden} if we look at one-party subsystems only. For
the 5-party case the number of non-local parameters that cannot be
seen by looking at each party locally only is $\leq
58$. There are at least $18$ non-local parameters that cannot be accessed by looking at any one and two-party subdensity matrices of the system.\newline
 In general the number of {\em hidden nonlocalities} grows exponentially with the number of parties.

\section{A class of $3$-LOCC incommensurate states}

We now want to show that there are $3$-LOCC incommensurate states
for the $3$-spin-1/2 system.  There are $5$ independent invariants under
local unitaries, $3$ of them are of the form $tr\rho _p^{2}$ ($p=A,B,C$) and completely characterize the eigenvalues of $\rho _p$ (see Section \ref{incom}). Suppose that we have two states $|\Psi
\rangle $ and $|\Phi \rangle $ that differ only in the last two {\em hidden}
invariants, i.e. the three one-party density-matrices of $|\Psi
\rangle $ and $|\Phi \rangle $ have the same eigenvalues. For the ease of argument choose them such that $\rho _A$, $\rho_B$
and $\rho_C$ have full rank $2$.\newline {\bf Claim:} {\em $|\Psi \rangle $
and $|\Phi \rangle $ are $3$-LOCC incommensurate.}\newline
{\bf Proof:} How does a general $3$-LOCC protocol look like?
First one party, say Alice, will perform a generalized measurement
and broadcast her outcome. Then Bob and Charlie will continue with
generalized measurements on their subsystems conditional on
Alice's outcome and broadcast their outcomes. At a certain point
Alice will continue and so on. Let's for a moment (mentally) merge
Bob's and Charlie's systems and look at the $3$-LOCC protocol as a
protocol between the systems A and BC. Everything that Bob and
Charlie do after receiving Alice's outcome and before Alice's next
action can be viewed as a generalized measurement on the BC
subsystem. So the whole $3$-LOCC protocol can be viewed as a
specific case of a $2$-LOCC protocol between A and BC. In particular
this means that if $|\Psi\rangle$ could be transformed into
$|\Phi\rangle$ via $3$-LOCC it certainly could be transformed into
$|\Phi\rangle$ via $2$-LOCC on A and BC.\newline Assume there were a
$3$-LOCC protocol that transforms $|\Phi\rangle$ to $|\Psi\rangle$.
We have chosen the states such that
\begin{equation}
\rho _{A}^{\Psi }\sim \rho _{A}^{\Phi }\sim \left(
\begin{array}{cc} \cos^{2} \alpha  & 0 \\ 0 & \sin^{2} \alpha
\end{array} \right).
\end{equation}
Let $|v_{A}\rangle $ and $|v_{A}^{\perp }\rangle $ be the two
eigenvectors of $\rho _{A}^{\Psi }$ and rewrite
\begin{equation}
|\Psi \rangle =cos\alpha |v_{A}\rangle |v_{BC}\rangle +sin\alpha
|v_{A}^{\perp }\rangle |v_{BC}^{\perp }\rangle
\end{equation}
This is the Schmidt decomposition of $|\Psi \rangle $ as a
bipartite A-BC state. In particular $|v_{BC}\rangle $ and $|v_{BC}^{\perp
}\rangle $ are orthogonal in the joint BC-Hilbert space.\newline
Alice performs the first generalized measurement
$M=\{M_{1},M_{2},\ldots \}$ (with $\sum_{i}M_{i}^{\dagger
}M_{i}=I$) and obtains the outcome $i$. Hereby she transforms the
state $|\Psi\rangle$ to a state $|\Psi ^{\prime }\rangle $ with
\begin{equation}
|\Psi ^{\prime }\rangle =\frac{1}{N}(cos\alpha (M_{i}|v_{A}\rangle
)|v_{BC}\rangle +sin\alpha (M_{i}|v_{A}^{\perp }\rangle
)|v_{BC}^{\perp }\rangle ).
\end{equation}
($N$ is the normalization factor.)\newline From here Bob and
Charlie will continue $|\Psi ^{\prime }\rangle \stackrel{
3-\text{LOCC}}{\longrightarrow }|\Phi\rangle$ so in particular we know
that $|\Psi ^{\prime }\rangle
\stackrel{2-\text{LOCC}\,\text{on}\,A,BC}{\longrightarrow }|\Phi\rangle$. From
\begin{equation}
\label{2loccprot}
|\Psi\rangle\stackrel{2-\text{LOCC}}{\longrightarrow }|\Psi ^{\prime
}\rangle \stackrel{2-\text{LOCC} }{\longrightarrow }|\Phi\rangle
\end{equation}
Nielsen's criterion (\ref{majo}) tells us
\begin{equation}
\rho _{A}^{\Psi }\prec \rho _{A}^{\Psi ^{\prime }}\prec \rho
_{A}^{\Phi }\sim \rho _{A}^{\Psi }.
\end{equation}
so $\rho _{A}^{\Psi }=cos^{2}\alpha |v\rangle \langle
v|+sin^{2}\alpha |v^{\perp }\rangle \langle v^{\perp }|$ has to
have the same eigenvalues as
\begin{eqnarray}
\rho _{A}^{\Psi ^{\prime }}&=&\frac{1}{N^{2}}(cos^{2}\alpha
M_{i}|v\rangle \langle v|M_{i}^{\dagger }+sin^{2}\alpha
M_{i}|v^{\perp }\rangle \langle v^{\perp }|M_{i}^{\dagger
})\nonumber \\ &=&\frac{M_{i}}{N}\rho _{A}^{\Psi }\frac{
M_{i}^{\dagger }}{N}.
\end{eqnarray}
This implies that there is a unitary transformation $U$ s.t.
\begin{equation}
U^{\dagger }\rho _{A}^{\Psi }U=\frac{M_{i}}{N}\rho _{A}^{\Psi
}(\frac{M_{i}}{ N})^{\dagger }  \label{eq:foutra}
\end{equation}
It follows that $U\frac{M_{i}}{N}$ has to be unitary and diagonal
in the same basis as $\rho _{A}^{\Psi }$ ( just pick a basis where
$\rho _{A}^{\Psi }$\ is diagonal and write out the matrix-elements
in their most general form
using that $\rho _{A}^{\Psi }$ has full rank). We see that $M_{i}|v_{A}\rangle $ and $%
M_{i}|v_{A}^{\perp }\rangle $ have to be orthogonal and that
Alice's generalized measurement reduces to a local unitary
operation on her qubit. \newline Continuing this argument for each
subsequent step of the $3$-LOCC protocol it follows that the whole
protocol ends up to be a succession of local unitaries. \newline
But we have chosen our states to be non-equivalent under local
unitaries. This completes the proof.  \newline

Note that the constraint to full-rank local density matrices can
be lifted if we only look at the restriction of $M_i$ onto the
support of $\rho_A$, $\rho_B$, $\rho_C$.\newline

We have thus shown that even in the simplest $3$-party case there
are states that--having the same eigenvalues of all
sub-density-matrices--are $3$-LOCC-incommensurate.
Furthermore once we fix the eigenvalues of $\rho_A$, $\rho_B$, $\rho_C$ we
have two additional parameters to specify different classes of
$3$-LOCC-incommensurate states. In the $5$-dimensional space of unitarily
non-equivalent states we have found a $2$-dimensional subspace of
mutually incommensurate states. \newline
This proof generalizes trivially to more than $2$
dimensions of each party's Hilbert space and  to $k \geq 3$ parties. To see the latter we note that at each step of a $k$-LOCC protocol we can divide the system into two parts--one party that performs a local operation and the other $k-1$ parties--and apply Nielsen's criterion as in the $3$-party case.\\
It follows from our proof that throughout each step of a $k$-LOCC
transformation protocol each party's density matrices of the state
obtained at a particular step have to majorize the corresponding
density matrices of all states at previous steps. In particular we have the \newline {\bf Corollary 1:}
If--say--Alice's density matrix at the beginning and at the end of
a $k$-LOCC protocol are similar then Alice's action is restricted to
local unitaries.\newline
{\bf Corollary 2:} Two $k$-partite states $|\Psi\rangle$ and $|\Phi\rangle$ that have similar density matrices ($\rho^{\Psi}_p \sim \rho^{\Phi}_p$, $p=A, B, \ldots$) on each party's subsystem are {\em $k$-LOCC incommensurate} if and only if they are {\em not unitarily equivalent}. \newline

After having proved the existence of $3$-LOCC incommensurate states
let's give some specific examples:

\subsection{The $2$GHZ-$3$EPR example}
\label{EPRGHZ}

BPRST$^1$ \cite{BPRST}, \cite{Popescupc}  originally showed that the following $3$-partite states are
incommensurate:
\begin{eqnarray}
2\text{GHZ}&=&\frac{(|0_{A1}0_{B1}0_{C1}\rangle +|1_{A1}1_{B1}1_{C1}\rangle )}{\sqrt{2%
}} \nonumber \\ &\otimes& \frac{(|0_{A2}0_{B2}0_{C2}\rangle +|1_{A2}1_{B2}1_{C2}\rangle )}{%
\sqrt{2}}  \label{eq:2GHZ}
\end{eqnarray}
and
\begin{eqnarray}
3\text{EPR}&=&\frac{(|0_{A1}0_{B1}\rangle +|1_{A1}1_{B1}\rangle
)}{\sqrt{2}}\nonumber \\ &\otimes&
\frac{(|0_{A2}0_{C1}\rangle +|1_{A2}1_{C1}\rangle )}{\sqrt{2}} \nonumber \\ &\otimes & \frac{%
(|0_{B2}0_{C2}\rangle +|1_{B2}1_{C2}\rangle )}{\sqrt{2}}  \label{3EPR}
\end{eqnarray}
In the $3$EPR-state the three parties Alice, Bob and Charlie share three
EPR-pairs, one between A and B, one between A and C and one between B and C.
In the $2$GHZ state they share just share two GHZ-states. In both cases the
density matrices of Alice, Bob and Charlie are identical:
\begin{equation}
\rho =\frac{1}{4}I  \label{eq:rho1}
\end{equation}
So in any LOCC transformation protocol from $2$GHZ to $3$EPR and vice
versa Alice, Bob and Charlie are restricted to local unitaries. It
is however impossible to transform $2$GHZ to $3$EPR via local
unitaries. One simple way to see this is to observe that $2$GHZ is a
tri-separable state--gives separable density matrices when
tracing out any one party--whereas $3$EPR is not tri-separable.
(Tracing out $A$ in $3$EPR gives $\frac{1}{4}I\otimes
|\text{EPR}_{B2C2}\rangle \langle \text{EPR}_{B2C2}|$ which is obviously not
separable.)\newline Note that this proof generalizes trivially to
$k$-partite states and $k$-LOCC:
\newline
{\em $(k-1)\text{GHZ}$ and ${k}\choose{2}$ \text{EPR} are $k$-LOCC
incommensurate.}

\subsection{Two locally non-distinguishable $3$-LOCC incommensurate
states of dimension 8} \label{incom}

Note that the smallest bipartite system that contains two
incommensurate states has to have dimension $9$ at least with each
party possessing a qutrit. This is because two density matrices
that are not majorized either way have to have rank $3$ at least.
But even for the smallest three-partite system of dimension $8$
there are incommensurate states. We can find states with identical
local density matrices that cannot be transformed into each other
via $3$-LOCC. To keep calculations easier we looked for particularly
simple states of the following form:
\begin{equation}
\label{simple}
 |\Psi \rangle =\alpha _{+}|000\rangle +\alpha
_{-}|vvv\rangle
\end{equation}
where $|v\rangle $ is a normalized state. The equivalence classes
of these states are characterized by two parameters--say $\alpha
_{+}\alpha _{+}^{*}$ and $|\langle 0|v\rangle |$--and have
equivalent density matrices on all three sub-parties. So from the $5$
independent invariants of (generic) states under local unitary
transforms in this case only (at most) $2$ are algebraically
independent.\newline Let's look at the invariants in the general
case for states of the form $ |\Psi \rangle =\sum_{i,j,k}\alpha
_{ijk}|e_{i}e_{j}e_{k}\rangle $. From the coefficients $\alpha
_{ijk}$\ we can form polynomials that are manifestly invariant
under local unitaries, like the degree $2$ polynomial:
\begin{equation}
I_{1}=\sum_{ijk}\alpha _{ijk}\alpha _{ijk}^{\ast }
\end{equation}
which is the norm of the state. To fourth degree we get three
polynomials:
\begin{eqnarray}
I_{2}&=&\sum_{ijkmpq}\alpha _{kij}\alpha _{mij}^{\ast }\alpha
_{mpq}\alpha _{kpq}^{\ast }=tr\rho _{A}^{2}\nonumber \\
I_{3}&=&\sum_{ijkmpq}\alpha _{ikj}\alpha _{imj}^{\ast }\alpha
_{pmq}\alpha _{pkq}^{\ast }=tr\rho _{B}^{2}\nonumber \\
I_{4}&=&\sum_{ijkmpq}\alpha _{ijk}\alpha _{ijm}^{\ast }\alpha
_{pqm}\alpha _{pqk}^{\ast }=tr\rho _{C}^{2}
\end{eqnarray}
which in general are algebraically independent. One of the higher
degree invariants is
\begin{equation}
\label{ifive}
I_{5}=\sum_{ijklmnopq}\alpha _{ijk}\alpha _{ilm}^{\ast }\alpha _{nlo}\alpha
_{pjo}^{\ast }\alpha _{pqm}\alpha _{nqk}^{\ast }
\end{equation}
which in general is not algebraically dependent of $I_{2},I_{3},I_{4}$. For
the simple state above (\ref{simple}) we have $I_{2}=I_{3}=I_{4}$
and a symbolical calculation (Groebner Basis) shows that $I_{5}$ and
$I_{2}$ are algebraically independent. We now exhibit two states of the above
simple form (\ref{simple}) which have similar one-party density
matrices (and the same $I_{2}$ ) but different $I_{5}$ thus being
{\em $3$-LOCC incommensurate}:

\begin{equation}
|\Psi \rangle =2\sqrt{\frac{3}{37}}|000\rangle -\frac{5}{\sqrt{37}}%
|111\rangle
\end{equation}
and
\begin{equation}
|\Phi \rangle =4\sqrt{\frac{2}{37}}|000\rangle -\frac{5}{\sqrt{37}}%
|vvv\rangle
\end{equation}
where $|v\rangle =\frac{|0\rangle +|1\rangle }{\sqrt{2}}$\ is the state $
|1\rangle $ rotated by 45 degrees.
\begin{equation}
I_{2}^{\Psi }=I_{2}^{\Phi }=\frac{769}{1369}
\end{equation}
and
\begin{equation}
I_{5}^{\Psi }\approx 0.343\neq I_{5}^{\Phi }\approx 0.242
\end{equation}
So these two states are $3$-LOCC incommensurate. We can apply a
local unitary transformation on each sub-system to one of the
states to make their density-matrices diagonal in the same basis
so that they are completely indistinguishable for each party.

\subsection{The $((3,2))$-threshold states}
\label{32ts}

In a recent paper \cite{CGL} an encoding of a {\em qutrit} into a tripartite state has been given (see Section \ref{TS}). The encoded state is of the following form:
\begin{eqnarray}
\label{32}
|\Phi(\alpha,\beta,\gamma)\rangle &=& \alpha (|000\rangle + |111\rangle + |222\rangle )+ \nonumber \\ && \beta (|012\rangle + |120\rangle + |201\rangle )+ \nonumber  \\ && \gamma  (|021\rangle + |102\rangle + |210\rangle )
\end{eqnarray}
The density matrix of any one party is proportional to the identity matrix. So all of these states have the same one-party density matrices. Most of these states will differ in the hidden nonlocalities and be $3$-LOCC incommensurate. Here we will give a set of three locally indistinguishable {\em orthogonal} states of the above form: 
\begin{eqnarray}
\label{32o}
|\Phi_1\rangle &=& |\Phi(1,0,0)\rangle = (|000\rangle + |111\rangle + |222\rangle ) \nonumber \\ |\Phi_2\rangle &=& |\Phi(0,\frac{1}{\sqrt{2}},\frac{1}{\sqrt{2}})=\frac{1}{\sqrt{2}} (|012\rangle + |120\rangle + |201\rangle ) \nonumber  \\ && +\frac{1}{\sqrt{2}}  (|021\rangle + |102\rangle + |210\rangle ) \nonumber \\ |\Phi_3\rangle &=& |\Phi(0,\frac{1}{\sqrt{2}},-\frac{1}{\sqrt{2}})\rangle = \frac{1}{\sqrt{2}}(|012\rangle + |120\rangle + |201\rangle ) \nonumber  \\ && -\frac{1}{\sqrt{2}} (|021\rangle + |102\rangle + |210\rangle )
\end{eqnarray}
These states differ in the value of $I_5$ (\ref{ifive}) which takes on 1/9, 1/18 and 0 for the three states respectively. They are thus $3$-LOCC incommensurate.

\section{Cryptography--Encoding into Hidden Nonlocalities}

We have exhibited states that cannot be transformed into each
other by local operations and classical communication involving
three parties and shown that there is a large number of them. We
think that these states can have a fruitful application in
(quantum)-cryptographic protocols like $3$-party quantum bit
commitment schemes. We can produce states that for each subsystem
are indistinguishable and yet have some hidden non-local property
that makes them different.\newline How could we {\em encode
information into the hidden non-localities} and how can we access
them?  One possibility is to find {\em orthogonal states} that
have the same respective sub-party density matrices and differ
only in these hidden parameters. Encode a bit-string into each of
those and give a part of the corresponding state to the three
parties, A, B and C without telling them which specific state they
share. While the parties are locally separated and only allowed to
perform local actions and classical communication  they have no
way to transform the states into each other. Only when they get
together (or send their share of the state through a quantum
channel) they can perform an orthogonal measurement and determine
the encoded bitstring. To ensure that there is no common state $\Omega$ from which {\em two different} states can be obtained via $k$-LOCC we can choose states with local density matrices proportional to the identity.  Since the identity
matrix on a subsystem majorizes every other density matrix,
$\rho_A^{\Omega}$ would have to be proportional to the identity as
well. We have shown that in this case each party is restricted to local
unitaries in their attempt to change the state via LOCC. \newline Another example of $2$ tri-partite states (apart from (\ref{32o}))
that {\em have identical one-party density matrices, are
$3$-LOCC-incommensurate and orthogonal} is the actual $2$GHZ-$3$EPR
example from Section \ref{EPRGHZ} if we use the singlet state
\begin{equation}
\text{EPR}^{^{\prime }}=\frac{1}{\sqrt{2}}(|00\rangle -|11\rangle)
\end{equation}
instead of the EPR states. $3$EPR' and $2$GHZ are $3$-LOCC incommensurate and
orthogonal.\newline
A detailed analysis of some cryptographic schemes for the potential use of incommensurate states should be done. Here we will restrict ourselves to a rather illustrative example involving quantum secret sharing.

\subsection{How Bob can cheat using the $((3,2))$ threshold scheme and how to prevent that}
\label{TS}
The $((3,2))$ threshold scheme in \cite{CGL} encodes a qutrit 
\begin{equation}
\label{qutrit}
|\Psi\rangle=\alpha|1\rangle+\beta|2\rangle+\gamma|3\rangle 
\end{equation} 
into the state $|\Phi(\alpha,\beta,\gamma)\rangle$ (\ref{32}). Each party obtains one qutrit of the encoded state: Alice the first, Bob the second and Charlie the third. This scheme allows any two parties together to completely extract the secret state $|\Psi\rangle$ (\ref{qutrit}). But no party alone can infer {\em any} information about the secret state $|\Psi\rangle$: each party's local density matrix is proportional to the identity.\newline
The procedure to retrieve $|\Psi\rangle$ from say the first two qutrits is the following \cite{CGL}: First the first register is added to the second (modulo 3) and then the (resulting) second qutrit is added to the first (mod 3). These operations can be performed without any measurement. This changes an encoded state $|\Phi(\alpha,\beta,\gamma)$ to
\begin{eqnarray}
\label{decode}
&&\begin{array}{c} \alpha (|000\rangle+|111\rangle+|222\rangle)+\\ \beta (|012\rangle+|120\rangle+|201\rangle)+ \\ \gamma (|021\rangle+|102\rangle+|210\rangle) \end{array} \stackrel{AB}{\longrightarrow} \begin{array}{c} \alpha (|000\rangle+|021\rangle+|012\rangle)+\\ \beta (|112\rangle+|100\rangle+|121\rangle)+\\ \gamma (|221\rangle+|212\rangle+|200\rangle) \end{array}\nonumber \\
&&=(\alpha|0\rangle+\beta|1\rangle+\gamma|2\rangle)\otimes(|00\rangle+|21\rangle+|12\rangle)
\end{eqnarray}
The secret state is completely restored in the first register. Analogous decoding procedures apply for AC and BC. \newline
In the original scheme \cite{CGL} it is assumed that the parties are honest when they participate in reconstructing the secret quantum state. \newline
Now assume the president of the bank uses this procedure to encode one of three (classical) ``trits'' ($b=0,1$ or $2$). He may want to use three orthogonal states $|\Phi_{0}\rangle$, $|\Phi_1 \rangle$, $|\Phi_2 \rangle$ that can be completely distinguished by an orthogonal measurement. So he distributes one of three {\em known} orthogonal states to his three vice-presidents. He does not want any of them {\em alone} to get knowledge about the encoded trit, only two of them together should be able to find out what the secret was. For illustration let's suppose that he uses the states
\begin{equation}
|\Psi_0\rangle=|0\rangle \;\;\;|\Psi_1\rangle=|1\rangle\;\;\;|\Psi_2\rangle=|2\rangle
\end{equation}
to encode $b=0,1$ and $2$ respectively and creates and distributes one of the three encoded states $|\Phi(1,0,0)\rangle$, $|\Phi(0,1,0)\rangle$ resp. $|\Phi(0,0,1)\rangle$. The three parties know the set of encoded states but not the actual state they are sharing. \newline
Now let's assume Bob decides to obtain the secret on his own without having to share his knowledge with Alice or Charlie. He thinks of the following strategy: \newline
He applies a unitary transformation $U$ to his share of the encoded secret state\begin{equation}
U:\;|0\rangle \rightarrow |1\rangle\;\;|1\rangle \rightarrow |2\rangle\;\;|2\rangle \rightarrow |0\rangle
\end{equation}
The set of encoded states after this transformation will have changed to 
\begin{eqnarray}
|\Phi(1,0,0)\rangle &\rightarrow& (|010\rangle+|121\rangle+|202\rangle)\nonumber\\|\Phi(0,1,0)\rangle &\rightarrow& (|022\rangle+|100\rangle+|211\rangle)\nonumber\\|\Phi(0,0,1)\rangle &\rightarrow& (|001\rangle+|112\rangle+|220\rangle
\end{eqnarray}
Alice and Charlie have no way of detecting Bob's dishonest action.
Suppose now that at the time for two parties to find out what the
secret was, Alice and Bob were the two to jointly retrieve the state. If they apply (\ref{decode}) to the changed state they obtain 
\begin{equation}
\begin{array}{cc }b=0:& |1\rangle\\
b=1:& |2\rangle\\ b=2:& |0\rangle \end{array}\;\;\;\;\otimes(|10\rangle+|01\rangle+|22\rangle) 
\end{equation}
At the end of this procedure Alice and Bob are supposed to know the
value of $b$. Assume $b=0$. Alice will think that $b=1$. Bob, however,
having changed the state, knows that if he jointly with Alice gets the
outcome ``$b=1$'', the actual trit $b$ is $0$! So he has obtained the
actual secret alone and misled Alice! Bob can apply $U^{-1}$ afterwards to erase the traces of his cheating completely. Similar misleading happens if
Bob and Charlie retrieve the secret. Of course if Alice and Charlie were the two to recover the secret trit Bob's action would not help and they will obtain $b=0$.\newline 
This type of cheating is possible, because the set of orthogonal states chosen to encode $b$ is equivalent under local unitaries. Bob has applied a local unitary transformation $U$ to change $\rho_{AB}^0 \rightarrow \rho_{AB}^1$ etc. Hereby he has changed the state corresponding to $b=0$ to a state $|\Psi\rangle$ with $\rho_{AB}^{\Psi}=\rho_{AB}^1$. $|\Psi\rangle$ and the actual state corresponding to $b=1$ are related by a local unitary on Charlie's system ($|\Psi\rangle$ is a purification of $\rho_{AB}^1$ and so is the state corresponding to $b=1$). This can only be possible if the states encoding $b=0$ and $b=1$ are unitarily equivalent.\newline
To prevent this type of cheating by a dishonest misleading party, the president of the bank has to select a set of {\em incommensurate} orthogonal states like (\ref{32o}). They are not transformable into each other by any local action (and classical communication). The class of incommensurate states has helped us to {\em choose a quantum secret}.

\section{Conclusion}

We have exhibited a class of locally equivalent multipartite
states that belong to an essential different class of
entanglement. Actually {\em almost all} locally similar
multi-particle states cannot be transformed into each other either
way by local operations and classical communication; {\em they are
  incommensurate}. The partial order induced on multipartite states by
transformation via $k$-LOCC is different from the bipartite case:
There is a multidimensional manifold of unitarily nonequivalent states
that are maximal in the sense that there is no other state from which they can be obtained by $k$-LOCC. The number of
parameters to characterize different classes of entanglement grows exponentially with the
number of parties involved. This space of locally
indistinguishable and yet incommensurate states suggests itself
for cryptographic applications involving several parties. We have shown that a set of incommensurate orthogonal and locally indistinguishable states can improve an $((n,k))$-threshold scheme against a form of cheating by a party. \newline
Other possible applications in cryptography should be
investigated. For instance, it is conceivable to find states $|\Omega
\rangle$ shared between $k$ parties such that any of them by
choosing a local action could transform the whole state into
either $|\Phi\rangle$ or $|\Psi\rangle$ where the last two states
have the same local density matrices for each party. This shared
state can then be used to share a secret between multiple users
that none of them can reveal to an outsider. Only in getting
together they can find out what the secret was.\newline The partial order of multipartite states should be investigated beyond classes of locally equivalent states.\newline
Another way to follow would be to suggest ``multipartite'' quantum bit commitment schemes involving sets of incommensurate states. Note that all proofs of the ``no-go'' theorem  \cite{Mayers}, \cite{LoChau1} for two-party quantum bit commitment schemes (like \cite{BCJL}) use the Schmidt-decomposition of a bipartite state (or in other words the non-existence of hidden parameters for two-party entanglement!). Multiparty protocols do not obey their line of argument.

\section{Acknowledgments}

Above all I would like to thank Michael Nielson for introducing me
into the area of multipartite entanglement, very fruitful discussions
and encouragement. Thanks to John Preskill for very helpful
discussions and hospitality at Caltech, where this work was done, and
to Alexei Kitaev for critical comments. Thanks also to Daniel Lidar
and Markus Grassl for improvements on the manuscript. \newline I wish to acknowledge continuing support by Issac Chuang. This
work was partially supported by DARPA DAAG 55-97-1-0341 and through the Quantum Information and Computing Institute (QUIC) administered through the ARO.

\end{document}